\documentclass[12pt]{article}
\usepackage{geometry}
\usepackage[utf8]{inputenc}
\usepackage{longtable} 
\usepackage{pdflscape} 
\usepackage{array} 

\usepackage{xcolor}
\definecolor{red}{rgb}{1, 0.0, 0.0}

\usepackage{amsmath}
\usepackage{amsfonts}
\usepackage{amssymb}
\usepackage{graphicx}
\usepackage{color}
\usepackage{enumitem}
\usepackage{eurosym}
\usepackage{wrapfig,lipsum,booktabs}
\usepackage{cyklop}
\usepackage{scrextend}
\usepackage{times}
\usepackage[colorlinks]{hyperref}
\usepackage{caption}
\usepackage{multirow}
\usepackage{xcolor,colortbl}
\usepackage{aas_macros}
\usepackage{natbib}

\usepackage{cancel}
\usepackage{ulem}
\usepackage{pifont}
\usepackage{amsthm}         
\usepackage{verbatim}
\usepackage{epsfig,bbm}
\usepackage{colortbl}
\usepackage{float}
\usepackage{subcaption}

\usepackage{epstopdf}
\usepackage{enumerate}
\input epsf
\usepackage{mathtools}
\usepackage{rotating}

\def\WMAP{\textrm{WMAP}}

\def\Planck{\textit{Planck}}
\def\muK{\ifmmode \,\mu$K$\else \,$\mu$\hbox{K}\fi}
\def\deg{\ifmmode^\circ\else$^\circ$\fi}
\def\think{Thi{\bf nK}}

\def\deg{\ifmmode^\circ\else$^\circ$\fi}
\def\pdeg{\ifmmode $\setbox0=\hbox{$^{\circ}$}\rlap{\hskip.11\wd0
.}$^{\circ}
          \else \setbox0=\hbox{$^{\circ}$}\rlap{\hskip.11\wd0
.}$^{\circ}$\fi}
\def\arcs{\ifmmode {^{\scriptstyle\prime\prime}}
          \else $^{\scriptstyle\prime\prime}$\fi}
\def\arcm{\ifmmode {^{\scriptstyle\prime}}
          \else $^{\scriptstyle\prime}$\fi}
\newdimen\sa  \newdimen\sb
\def\parcs{\sa=.07em \sb=.03em
     \ifmmode \hbox{\rlap{.}}^{\scriptstyle\prime\kern
-\sb\prime}\hbox{\kern -\sa}
     \else \rlap{.}$^{\scriptstyle\prime\kern -\sb\prime}$\kern -\sa\fi}
\def\parcm{\sa=.08em \sb=.03em
     \ifmmode \hbox{\rlap{.}\kern\sa}^{\scriptstyle\prime}\hbox{\kern-\sb}
     \else \rlap{.}\kern\sa$^{\scriptstyle\prime}$\kern-\sb\fi}

\providecommand{\sorthelp}[1]{}

\begin{document}

\fontfamily{times}\selectfont

\pagenumbering{gobble}
{\raggedright
{\bf \huge
Astro2020 APC White Paper \linebreak 

The need for better tools to design future CMB experiments}
 \linebreak
\normalsize

\textbf{\bf Thematic areas:} \linebreak
Primary Area: Cosmology and Fundamental Physics
\linebreak

\textbf{\bf Principal Author:} \linebreak
Name: Gra\c{c}a	 Rocha
 \linebreak						
Institution:  Jet Propulsion Laboratory (JPL), Caltech
 \linebreak
Email: graca.m.rocha@jpl.nasa.gov
 \linebreak
Phone:  (818)3930095
 \linebreak

\textbf{\bf Co-authors:} A.J. Banday$^{1}$, R. Belen Barreiro$^{2}$, Anthony Challinor$^{3,4,5}$, Krzysztof M, G\'{o}rski$^{6}$, Brandon Hensley$^{7}$, Tess Jaffe$^{8,9}$, Jeff Jewell$^{6}$, Brian Keating$^{10}$, Alan Kogut$^{11}$, Charles Lawrence$^{6}$, Georgia Panopoulou$^{12}$, Bruce Partridge$^{13}$, Tim Pearson$^{12}$, Joe Silk$^{14,15,16}$, Paul Steinhardt$^{17,18}$, Ingunn Wehus$^{19}$ \&
\\
Jamie Bock$^{6,12}$, Brendan Crill$^{6,12}$, Jacques Delabrouille$^{20,21}$, Olivier Dor\'{e}$^{6,12}$, Raul Fernandez-Cobos$^{2}$, Anna Ijjas$^{22}$, Reijo Keskitalo$^{23}$, Alexei Kritsuk$^{10}$, Anna Mangilli$^{1}$, Lorenzo Moncelsi$^{12}$, Steve Myers$^{24}$, Bryan Steinbach$^{12}$, Matthieu Tristram$^{25}$.

}

\vspace{0.4in}
\noindent
{\scriptsize
%
$^1$ IRAP, Université de Toulouse, CNRS, CNES, UPS, (Toulouse), France
\\   
%
$^2$ Instituto de F\'{\i}sica de Cantabria (CSIC-Universidad de Cantabria), Avda. de los Castros s/n, Santander, Spain
\\   
%
$^3$ Institute of Astronomy, University of Cambridge, Madingley Road, Cambridge CB3 0HA, UK
\\   
$^{4}$ Kavli Institute for Cosmology Cambridge, Madingley Road, Cambridge, CB3 0HA, UK
\\ 
$^{5}$ DAMTP, Centre for Mathematical Sciences, University of Cambridge, Wilberforce Road, Cambridge CB3 0WA, UK
\\ 
%
$^{6}$ Jet Propulsion Laboratory, California Institute of Technology, 4800 Oak Grove Drive, Pasadena, California, USA
\\   
%
$^{7}$ Department of Astrophysical Sciences, 4 Ivy Lane, Princeton University, Princeton, NJ 08544, USA
\\   
%
$^{8}$ CRESST, NASA Goddard Space Flight Center, Greenbelt, MD 20771, USA
\\ 
$^{9}$ Department of Astronomy, University of Maryland, College Park, MD, 20742, USA
\\   
%
%
$^{10}$ Physics Department and Center for Astrophysics and Space Sciences, University of California San Diego, CA 92093-0424, USA
\\   
%
$^{11}$ Code 665, NASA Goddard Space Flight Center, Greenbelt, MD 20771, USA
\\   
%
%
$^{12}$ California Institute of Technology, MC249-17, 1200 East California Boulevard, Pasadena, CA 91125, USA
\\   
%
$^{13}$ Haverford College Astronomy Department, 370 Lancaster Avenue, Haverford, Pennsylvania, U.S.A.
\\   
%
%
$^{14}$ Institut d’Astrophysique de Paris, Sorbonne Universit{\'e}s, UPMC Univ. Paris 06 et CNRS, UMR 7095, F-75014, Paris, France
\\ 
$^{15}$ Department of Physics \& Astronomy, The Johns Hopkins University, Baltimore, MD 21218, USA
\\   
$^{16}$ Beecroft Institute of Particle Astrophysics and Cosmology, Department of Physics, University of Oxford, Oxford OX1 3RH, UK
\\ 
%
$^{17}$ Department of Physics, Princeton University, Princeton, NJ 08544, USA
\\   
$^{18}$ Princeton Center for Theoretical Science, Princeton University, Princeton, NJ 08544, USA
\\ 
%
$^{19}$ Institute of Theoretical Astrophysics, University of Oslo, Blindern, Oslo, Norway
\\   
%
$^{20}$ Laboratoire Astroparticule et Cosmologie (APC), CNRS/IN2P3, 10, rue Alice Domon et L\'eonie Duquet, 75205 Paris Cedex 13, France.
\\   
$^{21}$ D\'epartement d'Astrophysique, CEA Saclay DSM/Irfu, 91191 Gif-sur-Yvette, France.
\\   
%
$^{22}$ Institute for Theory and Computation, Harvard-Smithsonian Center for Astrophysics, Harvard University, Cambridge, MA 02138, USA
\\   
%
$^{23}$ Computational Cosmology Center, Lawrence Berkeley National Laboratory, Berkeley, California, U.S.A.
\\   
%
$^{24}$ National Radio Astronomy Observatory, 1003 Lopezville Road, Socorro, NM 87801, USA
\\   
%
$^{25}$ LAL, Univ. Paris-Sud, CNRS/IN2P3, Universit Paris-Saclay, Orsay, France
\\   
}

\pagebreak

 \pagenumbering{arabic}


\section{Introduction}

This white paper addresses key challenges for the design of
next-decade Cosmic Microwave Background (CMB) experiments, and for
assessing their capability to extract cosmological information from
CMB polarization.  We focus here on the challenges posed by foreground
emission, CMB lensing, and instrumental systematics to detect the
signal that arises from gravitational waves sourced by inflation and
parameterized by $r$, at the level of $r \sim 10^{-3}$ or lower, as
proposed for future observational efforts. We argue that more accurate
and robust analysis and simulation tools are required for these
experiments to realize their promise.  We are optimistic that the
capability to simulate the joint impact of foregrounds, CMB lensing,
and systematics can be developed to the level necessary to support the
design of a space mission at $r \sim 10^{-4}$ in a few years.  We make
the case here for supporting such work.  Although ground-based efforts
present additional challenges (e.g., atmosphere, ground pickup), which
are not addressed here, they would also benefit from these improved
simulation capabilities.

The expected inflationary signal has peaks at low multipoles,
$\ell \sim 8$, and mid-multipoles $\ell \sim 80$. Measurements at low
$\ell$ require near full-sky coverage, for which foreground are
$\sim$100 times larger in amplitude than the signal at 75\,GHz (for
60\% sky). Although measurements at the mid-multipoles can be
initially conducted on smaller patches of the sky, for which the
foregrounds are $\sim$10 times larger than the signal, this comes with
a penalty on the cosmic variance error from the signal from lensing of
the CMB photons, brighter than target inflationary B-modes by a factor
of 3 (assuming $r \sim 10^{-3}$) or more.

Hence, improvements in $r$ between 1 and 2 orders of magnitude with
respect to the current upper limit $r < 0.07 $~($95\%$)
~\citep{Ade:2018gkx} will require improvements in foreground
separation, CMB de-lensing, and systematics by similar or greater
factors.  No existing or proposed experiment has demonstrated the
capability to address all of these issues simultaneously at the
required level.  Control of foregrounds and systematics is also
important for attaining cosmic-variance-limited measurements of the
optical depth to reionization, $\tau$. The large-scale $E$-modes that
encode most of the relevant information are below the foregrounds at
$\ell < 10$, and systematics need to be controlled on the largest
angular scales at a matching level.  Furthermore, foregrounds play an
important role in recovering the spectral distortions signals in the
CMB.\footnote{This whitepaper is a summary of the report from the KISS
  workshop \citep{Rocha2019} `Designing future CMB experiments', held
  on March 19--23, 2018, at Caltech, Pasadena, CA, USA.  The report
  (in prep.)
  will appear here:\\
  \url{http://kiss.caltech.edu/workshops/fCMB/fCMB.html}}


\section{\think -  the philosophy in going from scientific goals to the mission }
In designing future missions, the fundamental question is what
measurements must be made in order to address the science goals?  Few
missions address only a single goal, and hardware designed to do one
thing well invariably does many things.  So the question becomes what
measurements that we need to make drive the design of the hardware.
In effect, what measurements are the most difficult to make, or
require the greatest capability?

As we have discussed, determination of $r$, is particularly hard.  It
requires measurement of fluctuations in polarization on large angular
scales at extremely low levels, which \WMAP\ and \Planck\ have
demonstrated is the hardest of all anisotropy measurements to make,
with the added complication that we do not know in advance at what
level the fluctuations will appear.  There are many models of
inflation, and many predictions of the size of $r$ over many orders of
magnitude (see Figure~\ref{inflation-pico}).  Since we do not know the
level of gravitational-wave-induced $B$-mode polarization in the
reionization and recombination peaks, we simply cannot specify how
well we must measure the sky to extract most of the relevant
information.

In designing experiments, then, we suggest the following two
principles:

\begin{itemize}
\item {\bf Take big steps, but not too big.}  Steps into the unknown
  carry significant uncertainty.  A distant goal is more effectively
  reached in two steps, with learning and correction after the first
  incorporated into the second, than in one giant step.  The kind of
  experiments that we are discussing are expensive, whether on the
  ground or in space.  Small, merely incremental steps do not justify
  the cost.  But too big steps risk going astray.  It's a matter of
  judgment, but hardheaded ambition rather than untethered dreaming
  should prevail.
\item {\bf Understand the real limitations of the measurements.}  For
  CMB experiments, noise has always been a major issue, and so it will
  remain.  However, systematic errors (``systematics'') have been the
  limitation for many experiments, from early attempts to measure the
  Solar dipole to the 2014 claim that primordial $B$-mode fluctuations
  were measured, when the fluctuations were, in fact, mostly due to
  polarized Galactic dust emission.
\end{itemize}

The critical point is that the limitation of future experiments will
be some combination of foregrounds and systematics.  Since in
principle instrumental systematics can be reduced, but foregrounds
cannot, it is inevitable that foregrounds will set the ultimate limit
on how well the CMB can be measured.

It is straightforward to calculate how well $B$- and $E$-mode power
spectra must be measured to determine $r$ to a certain level.
Straightforward, although not entirely simple: it matters, for
example, whether one is determining an upper limit, or measuring a
constrained value.  But the complications are far greater in trying to
predict the effects of foregrounds and systematics.  These can be
estimated before launch, but they can only be known for sure from the
data themselves.  The hard -- very hard -- part is to separate them
from one another, so that something can be done about them.  Only if a
certain feature or pattern in the data can be traced to a particular
instrumental behavior, known before launch or discovered in flight,
and simulated with confidence, is it possible to say that it is
understood, and to do something about it.  Correlation is not good
enough.  Systematic effects can be degenerate, or nearly so, where
certain combinations of systematics mimic each other.  The further
along in the analysis, the more this is true.  At the power spectrum
level, separation of systematics from each other is essentially
impossible.  At the map level, the situation is better, because the
two-dimensional nature of the data provides many more clues.  Other
problems, obviously including time-dependent ones, can only be
adequately understood at the time-ordered-data level.

Accurate quantitative assessments will require high fidelity
simulations, but we can make some crude estimates.  With $\tau$ (the
optical depth to reionization) of about 0.06, the reionization bump
($2\le\ell\le12$) in the $B$-mode power spectrum would be at a level
of about 1\,nK$^2$ for $r\sim 10^{-4}$.  The rms brightness
temperature of $B$-mode fluctuations on a 40\arcm\ scale at about
200\,GHz is about 1\,nK as well, a factor of 20 or so below the level
of {\it synchrotron} fluctuations at that frequency over 70\% of the
sky, and a factor of $\sim 2000$ below the level of {\it dust}
fluctuations.  Errors in the final Planck maps, after a decade of
processing, were of order 1\muK\ on a 1\deg\ scale, and supported an
upper limit on $r$ of 0.08.  To reach $r = 0.0001$, one might
calculate that maps would have to be $\sqrt{800}$ better, or in the
30\,nK range.  Given that the combined effects of residual foregrounds
and systematics are likely to get harder to disentangle at lower
levels, it is not hard to believe that map errors at a level of
$\leq10$\,nK are about the largest that can be tolerated.  We
summarize this level by saying ``think nK'', or \think.


\section{The Science Cases}

\def\simpropto{\lower.2ex\hbox{$\; \buildrel \propto \over \sim \;$}}
\def\ltsim{\lower.5ex\hbox{$\; \buildrel < \over \sim \;$}}
\def\gtsim{\lower.5ex\hbox{$\; \buildrel > \over \sim \;$}}

\begin{wrapfigure}{r}{0.5\textwidth}
\centering 
\includegraphics[width=0.5\textwidth]{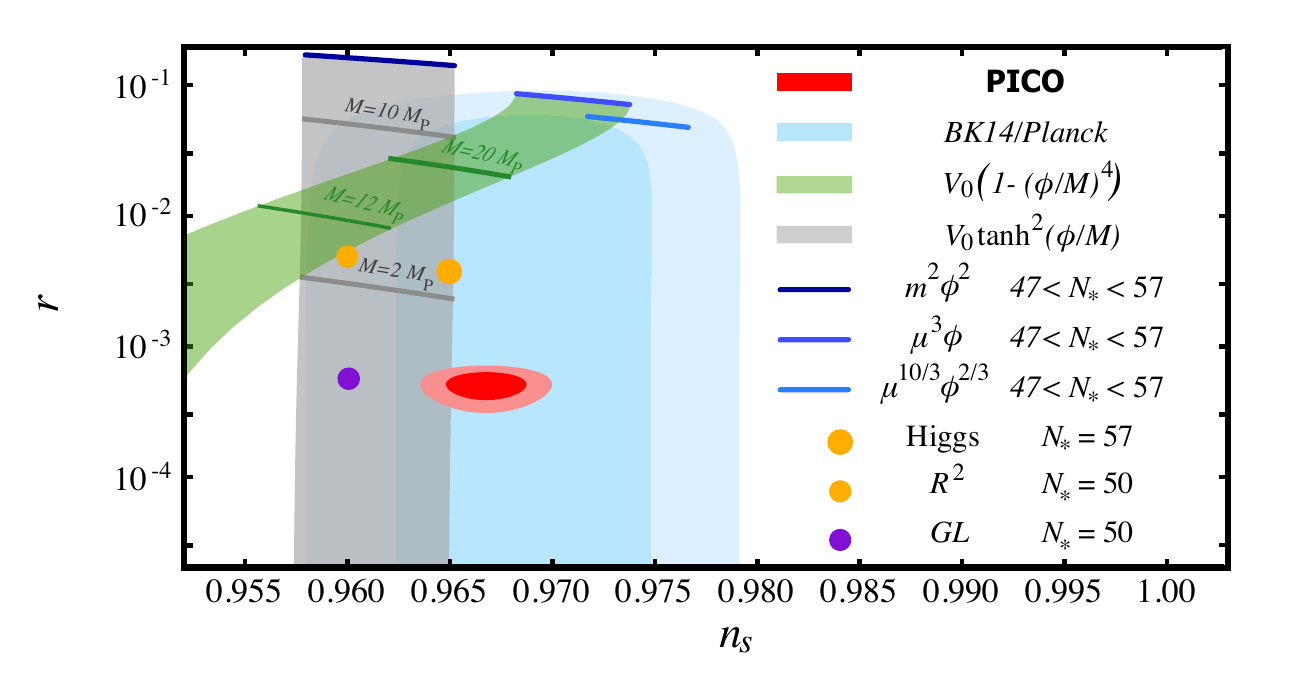} 
\caption{\label{inflation-pico} Current $1\sigma$ and $2 \sigma$ as limits on r and $n_{s}$
  (cyan) and forecasted constraints for a fiducial model with
  $r = 0.0005$ for PICO, together with predictions for selected models
  of inflation. Adapted from \cite{Hanany2019}.}
\end{wrapfigure}
The science case for future CMB experiments has been clearly laid out
in Astro2020 Science whitepapers~\citep[e.g.,][]{Shandera2019,Chluba2019}.
We list here a few key elements.
\newline
{\bf (a) Inflationary models} for the origin of cosmological
perturbations: These models postulate that the sources of
gravitational waves in the early universe are vacuum (quantum)
fluctuations from Inflation (an era of accelerated expansion in the
early universe).  These models have succeeded in resolving the
homogeneity, isotropy, flatness and monopole problems, and the puzzle
of explaining how the universe obtained a nearly scale-invariant
spectrum of density variations that extend to super-horizon scales.
Inflation requires exponentially fine-tuned initial conditions to
initiate a sufficiently long period of accelerated expansion.
Figure~\ref{inflation-pico} shows the predictions for the
tensor-to-scalar ratio, $r$, and the scalar spectral index, $n_{s}$,
of several Inflationary models along with constraints from Planck/BK2
and forecasts for PICO \citep{Hanany2019}.
\newline
{\bf (b) Non-inflationary models} for the origin of cosmological
perturbations: An example of non-Inflationary models are the Cyclic
models (see \citealt{Shandera2019} and references therein for other
models).  Cyclic models of the universe provide an explanation for the
homogeneity and flatness of the universe and the cosmic generation of
a nearly scale-invariant gaussian spectrum of density perturbations
while avoiding any kind of initial condition or multiverse problems.
Inflationary expansion is replaced by ekpyrotic (ultra-slow)
contraction and the big bang is replaced by a transition from
contraction to expansion, sometimes referred to as a ``bounce."  The
resulting cosmology not only resolves the standard cosmological
conundra, but also implies the {\it absence of B-modes,} evades the
cosmological singularity problem and resolves the entropy problem of
earlier cyclic models.  Inflationary and cyclic scenarios have
important distinctions.  For example, in an inflationary multiverse, a
Hubble-sized patch of space like we observe today might be spatially
flat, but it can just as well be open or closed. In the cyclic
scenario, there is no option: the patch must be spatially flat,
period. And the same applies to other features of both the
Inflationary and cyclic models.
\newline 
{\bf (c) CMB spectral distortions}:

\begin{wrapfigure}{r}{0.5\textwidth}
\centering 
\includegraphics[height=1.85in,width=0.5\textwidth]{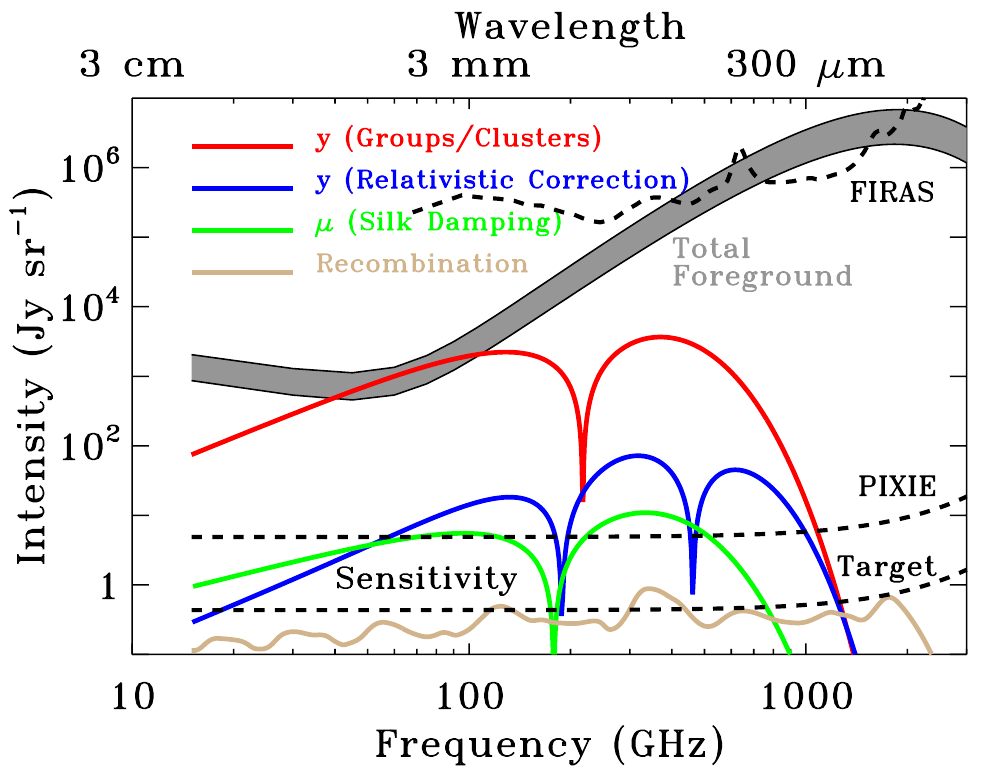} 
\caption{\label{SD2} Spectral distortions vs frequency, infrared foregrounds, the
  standard and relativistic $y$ and $\mu$ spectral distortions and the
  hydrogen and helium recombination lines; the proposed PIXIE
  sensitivity and the target sensitivity needed for a guaranteed
  minimal science return. Adapted from \cite{Desjacques2015}.}
\end{wrapfigure}
The sky-averaged CMB spectrum is known to be extremely close to a
perfect blackbody at a temperature \citep{Fixsen1996}
$T_0 = 2.7255\pm 0.0006$ K, with possible distortions limited to parts
in $10^5$.
Given the uncertain prospects for detecting a primordial B-mode signal
from inflation, it is relevant to consider complementary approaches
that are capable of yielding unique information on the primordial
universe, directly probing unprecedentedly early epochs back to the
epoch when the cosmic blackbody radiation originated.  Sources of
spectral distortion arise in the pre-recombination and
post-recombination epochs where spectral distortions emerge as a
combination \citep{Chluba2016} of late epoch $y$-distortions
\citep{Zeldovich1969} and early epoch $\mu$-distortions
\citep{Sunyaev1970} in the standard $\Lambda$CDM model.
Figure~\ref{SD2} shows these spectral distortions versus frequency and
the proposed PIXIE sensitivity \citep{Kogut2016, Desjacques2015}.


\section{A CMB Program: Space vs Ground Complementarity}
Historically, major steps in CMB observations have been made from the
ground, from balloons, and from space.  What of the future?  The
answer depends on the level of accuracy that must be achieved.

The advantages of space are full coverage of the frequency spectrum,
full coverage of the sky, freedom from systematic errors associated
with the Earth's atmosphere or surface, and stability.  The advantages
of the ground are accessibility and cost.

\begin{wrapfigure}{r}{0.5\textwidth}
\begin{centering}
\includegraphics[height=1.8in,width=0.5\textwidth]{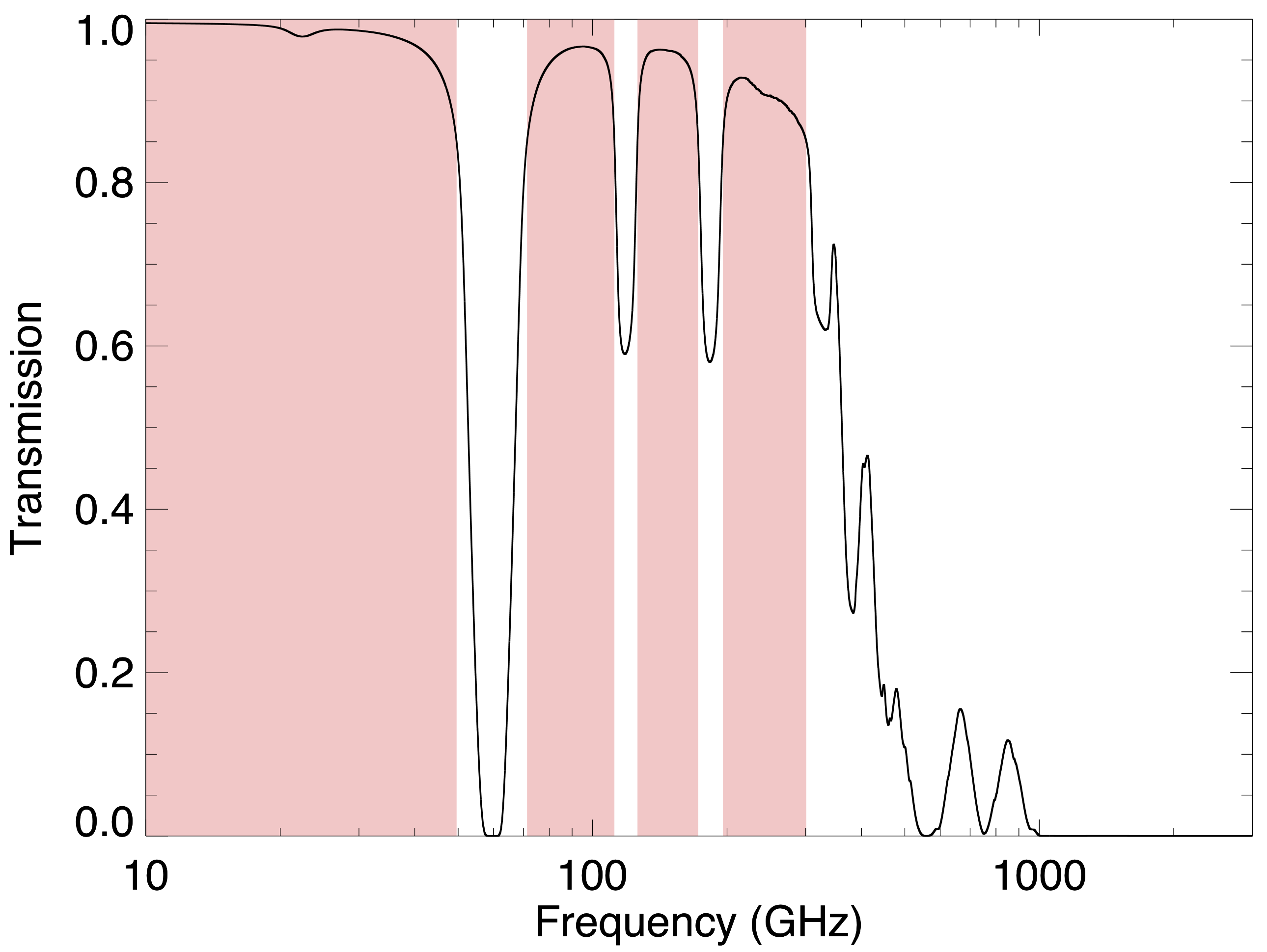}
\caption[Atmopsheric Windows] {\label{fig:atm} Atmospheric transmission vs frequency
  from a high, dry site (South Pole or the Atacama plateau).  Pink
  bands show the primary windows to observe the CMB. Measurements
  outside these windows are impossible from the ground.  }
\end{centering}
\end{wrapfigure}

Figure~\ref{fig:atm} shows typical atmospheric transmission from a
high, dry site such as the South Pole or the Atacama plateau in Chile.
Strong atmospheric absorption features limit ground-based observations
to frequencies below 45\,GHz or to relatively narrow ``windows''
centered at 90, 150, and 250\,GHz. Within these windows, atmospheric
emission is bright but largely unpolarized; its principal effect is
increased noise coupled with restrictions on effective scan
strategies.  To quantify the effect of the noise, the rule of thumb
among CMB experimentalists is that one detector in space is worth
100~detectors on the ground.  Moreover, comparison of
Figure~\ref{fig:atm} with Figure~\ref{fig:fgpol} shows that the
atmosphere is essentially opaque in the important minimum foreground
frequency band.

Space offers particular advantages for measurements on the largest
angular scales.  Free from diurnal temperature variation, space
platforms have demonstrated mK stability on time spans of months to
years, reducing effects of long-term calibration drifts.  With no
terrain to induce position-dependent ground pickup, offset drifts are
correspondingly minimized.  Orbital observatories have minimal
constraints on pointing pitch or roll angle, and can readily rotate
about the beam axis to generate full parallactic angle coverage within
each sky pixel.  These effects combine to control systematic errors,
which can otherwise dominate over instrument white noise on the
largest angular scales.

Space platforms are the only viable option to measure distortions from
the CMB blackbody spectrum ($\S$3).  Distortions from hot gas in
groups and clusters are expected at the 100\,nK level, with
distortions from the dissipation of primordial density perturbations
present at nK levels.  Separating these signals from competing
foreground emission requires continuous spectra calibrated to a common
standard across several octaves in frequency.  The resulting long
integration times outside the available atmospheric windows requires a
space mission.

On the other hand, the 5–10-m telescopes desirable for studying
neutrinos and secondary anisotropies are much more easily obtained on
the ground than in space.  Ground-based platforms are also more
flexible than space missions.  Multiple platforms operating at
multiple locations allow robust cross-checking of different detector
technologies and observing strategies.  Simple access to ground-based
observatories allows frequent incremental upgrades.  Ground-based
instruments develop and use cutting-edge technologies, while the
longer development cycle and lower risk tolerance for space missions
can leave such missions a decade behind ground-based state of the art.
The shorter development time and higher risk tolerance for
ground-based missions allows quick reaction to new discoveries.

To summarize, for $r \ge 10^{-3}$ from the recombination bump
($\ell > 30$), for neutrinos and secondary anisotropies, observations
from the ground may be possible and important.  For $r < 10^{-3}$,
especially for $\ell<30$, and for spectral measurements, space is the
only option.


\section{Systematics}
As instrumental sensitivities approach the nK threshold, instrument
design must become correspondingly robust against systematic effects
at that level.  Although instrumental effects are by their nature
instrument-specific, a robust design will include multiple lines of
defense:

\begin{itemize}
\item {\bf Eliminate:} Differential measurements are a powerful tool
  to eliminate broad classes of systematic errors.  By canceling
  common-mode signals, differential techniques prevent these signals
  from sourcing systematic error.  A common example is the
  differential comparison of signals from independent beams on the
  sky, which rejects emission from the CMB monopole and (for
  co-pointed beams) unpolarized anisotropy."  Beam ellipticity can
  couple to local gradients in the sky brightness to mimic sky
  polarization.  Differential measurements remove this common-mode
  signal component, eliminating temperature--polarization coupling to
  first order.
\item {\bf Mitigate:} Null measurements suppress systematic effects by
  reducing the amplitude of the underlying source terms.  For example,
  absorption and emission from optical surfaces within an instrument
  can impart instrumental polarization to modulate or mimic true sky
  polarization.  Instrumental polarization depends on the temperature
  difference between the sky and the optical surfaces within the
  instrument.  Maintaining the instrument within a few mK of the sky
  temperature suppresses instrumental polarization by 3--5 orders of
  magnitude compared to optics maintained at room temperature.
\item {\bf Modulate:} Modulation imparts a distinctive time-dependent
  signature on desired sky signals to distinguish them from
  instrumental effects that do not share the same time dependence.
  Polarization modulation is a common example.  A rotating half-wave
  plate placed as the first optical element within an instrument
  causes the plane of polarization from sky signals to rotate at twice
  the frequency of the wave plate.  Synchronous demodulation at twice
  the rotation frequency efficiently separates polarized sky signals
  from fixed instrumental polarization (constant in time) or even
  spurious signals from the rotator drive itself (typically occuring
  at the rotation frequency).  Sufficiently rapid modulation also
  mitigates slow drifts or $1/f$ noise, forcing the sky signal to
  frequencies above the $1/f$ knee.
\item {\bf Calculate:} Improvements in raw sensitivity require
  corresponding improvements in the modeling of instrumental effects.
  Reaching nK sensitivity and beyond demands calculation of effects
  beyond first order in perturbations.  To return to the example of
  beam ellipticity above, we see that differential beam subtraction
  eliminates temperature-polarization coupling from beam ellipticity
  to first order.  At second order, however, the {\it differential}
  beam ellipticity still couples with unpolarized gradients on the sky
  to mimic a polarized sky signal.  Similarly, the cross-polar
  response of the instrument beam pattern can couple with a
  transmissive half-wave plate to create $E \rightarrow B$
  polarization mixing, again at second order.  First-order suppression
  is not necessarily sufficient to reach nK sensitivity;
  next-generation CMB missions must calculate systematic error signals
  to higher order to ensure that errors are sufficiently suppressed.
\end{itemize}

As raw sensitivity improves, all possible techniques will be required
to reduce systematics to levels below the noise.  Presently available
tools are not reliable at the nK level required by the most ambitious
future experiments.  We identify development of improved tools as a
critical element in the CMB program, and ask for a recommendation of
support for such development.


\section{Foregrounds}

\begin{wrapfigure}{r}{0.5\textwidth}
\begin{centering}
\includegraphics[height=1.6in,width=0.45\textwidth]{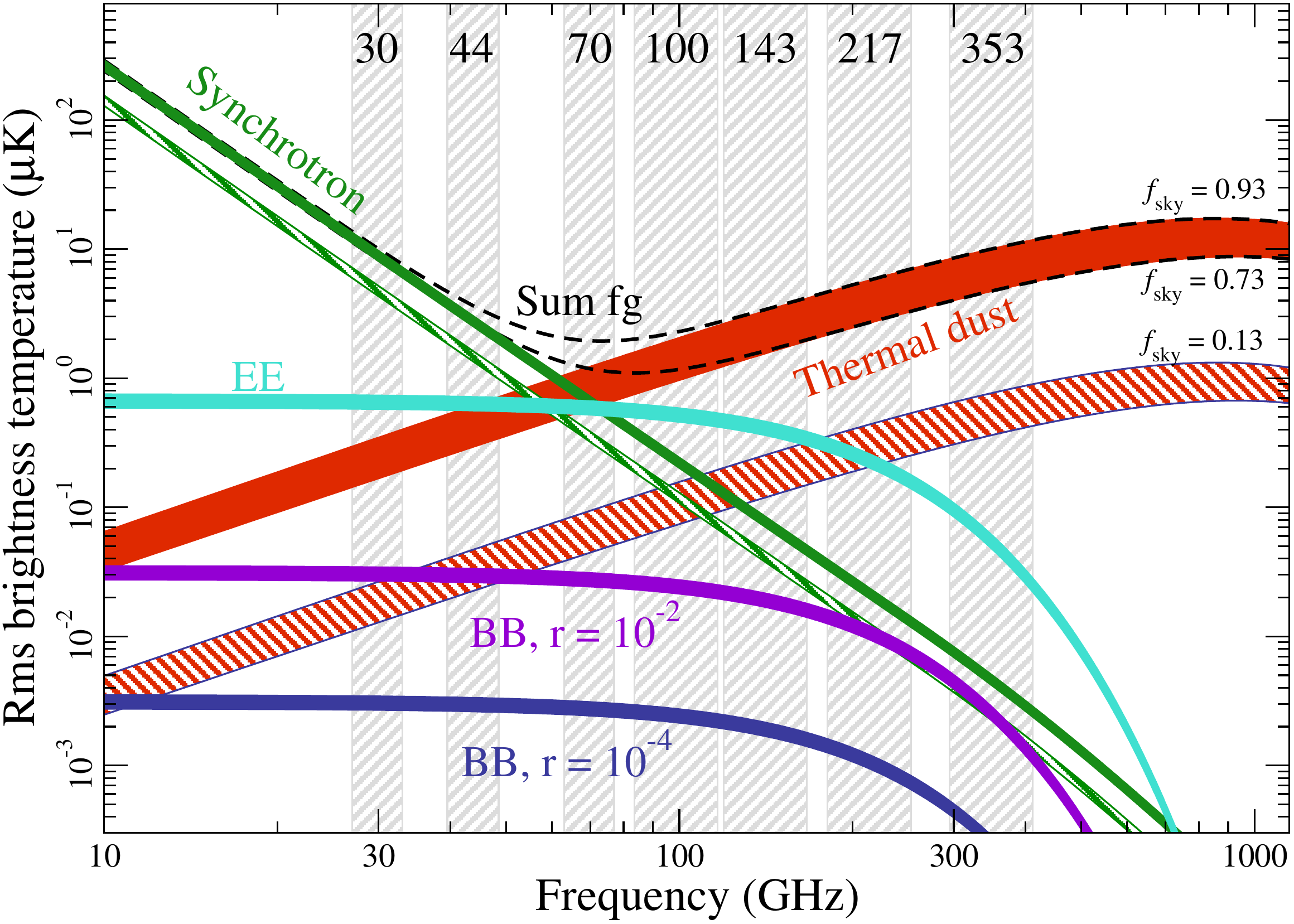}
\caption{\label{fig:fgpol} Brightness temperature rms as a function of frequency and
  astrophysical component for polarization. Based on figure 51 from
  \cite{Planck2015X}}
\end{centering}
\end{wrapfigure}

The Planck mission has demonstrated that Galactic emission will be the
dominant foreground, especially at large angular scales. The principal
diffuse Galactic foregrounds in polarization are dust emission, which
dominates at high frequencies, and synchrotron emission, which
dominates at low frequencies (see Figure~\ref{fig:fgpol}). However,
other diffuse foregrounds are known to contribute significantly in
total intensity and may be relevant in polarization as well,
particularly at the sensitivity levels of next generation missions.
These other emission mechanisms are: free-free emission, anomalous
microwave emission (AME), line emission from interstellar gas
(particularly the CO rotational lines), the Zodiacal Light at
far-infrared frequencies, and the Cosmic Infrared Background (CIB).

Known physical complexities in foreground emission not encapsulated by
simple parametric models, such as line of sight averaging and grain
alignment, should be quantified in terms of their effect on foreground
subtraction.  Models have been developed to describe the spatial
morphology and frequency dependence of many of these foregrounds,
enabling high fidelity component separation at current noise levels. A
key question for a future CMB mission is the extent to which the
parameterizations currently employed will describe the various
foregrounds at the accuracy demanded by future missions.  This
presents something of a conundrum--the presence of new subtleties in
foreground emission may drive choices in instrument design and data
analysis, but they can be truly characterized only by making the
measurements.
 
We postulate that there are three broad paths for bringing theory,
data, and simulations to bear on this problem and thereby improving
forecasting as well as informing the development of instrument designs
and component separation algorithms (\citealt{Rocha2019}).

\begin{itemize}
\item {\bf Connect Physical Foreground Models to Uncertainties in
    Foreground Subtraction}: The physics of the various emission
  mechanisms that constitute the CMB foregrounds is rich and
  complex. Known physical complexities in foreground emission not
  encapsulated by simple parametric models, such as line-of-sight
  averaging and grain alignment, should be quantified in terms of
  their effect on foreground subtraction. Detailed calculations should
  be made of the expected levels of polarization emission from
  free-free, CO, and Zodiacal Light. Empirical constraints on the
  polarization properties of these emission mechanisms should be
  established using existing data.
\item {\bf Develop More Sophisticated Simulations That Capture
    Important Known Complications of Foreground Emission}: The
  efficacy of foreground mitigation strategies, whether in instrument
  design or data analysis, should be assessed against simulations that
  reflect realistic levels of foreground complexity. Further
  development of existing sky models that use Galactic observations to
  constrain the spatial statistics of the various foregrounds (e.g.,
  the \Planck\ Sky Model, \citealt{Delabrouille2013}, and the Python Sky
  Model, \citealt{Thorne2017}) should be pursued. In particular,
  generating non-Gaussian realizations of foreground amplitudes at
  small multipoles would be valuable.  The use of Magnetohydrodynamic
  simulations, which can incorporate many of the relevant complexities
  in a natural, physically-motivated way, to generate mock microwave
  skies should be developed further
  \citep[e.g.,][]{Kritsuk2018,Kim2019}.
\item {\bf Explore How Existing and Upcoming Ancillary Datasets Can Be
    Incorporated in Component Separation}: The data landscape at the
  time of the next space mission will include detailed observations of
  the three dimensional, magnetized interstellar medium as well as
  catalogs of extragalactic sources at relevant
  frequencies. Assembling a comprehensive, multi-frequency picture of
  the 3D ISM is an active area of research, e.g., the IMAGINE
  Consortium \citep{imagine}
  \footnote{\url{https://www.astro.ru.nl/imagine/imagineprojects.html}}
  and Cosmoglobe\footnote{\url{http://cosmoglobe.uio.no}}, which will
  play an important role in informing and validating CMB component
  separation. Ancillary data from sub-orbital experiments can also
  play a crucial, direct role by extending frequency coverage to both
  higher and lower frequencies that cannot be realized on a single
  space mission.
\end{itemize}


\section{Lensing}
The CMB is gravitationally lensed by large-scale structure as it
propagates across the $14\,\text{Gpc}$ (comoving) distance from
recombination to the present. Lensing remaps the temperature and
polarized surface brightness, distorting our view of the primary CMB
fluctuations. This is both a blessing and a challenge. The distortions
due to lensing introduce very specific non-Gaussian statistics to the
lensed CMB, which can be measured to extract information on the
large-scale clustering of matter at intermediate redshifts that is
difficult to access by other means. However, lensing also blurs our
view of the infant universe. In particular, lensing converts $E$-mode
polarization into $B$-modes, confusing searches for the $B$-mode
signal from primordial gravitational waves expected from inflation.

Here, we focus on the issue of mitigating the effect of lensing on
searches for degree-scale primordial $B$-mode polarization. The
$B$-modes produced by lensing have an almost white-noise angular power
spectrum on large scales, equivalent to an additional map-level noise
of $5\,\mu\text{K}\,\text{arcmin}$. These have now been measured by
the BICEP/Keck Array experiments~\citep{Ade:2018gkx}, and on smaller
scales by SPTPol~\citep{Keisler:2015hfa}, ACTPol~\citep{Louis:2016ahn}
and POLARBEAR~\citep{Ade:2017uvt}.  If the tensor-to-scalar ratio
$r< 0.01$, the signal power is below that from lensing at multipoles
$l>10$; only the large-scale signal from reionization exceeds
lensing. However, the expected lensing power can already be predicted
at the percent level, and subtracting this from the measured power
allows one to access primordial gravitational waves with $r\ll
0.01$. Such subtraction does not remove the sample variance of the
lensing $B$-modes, though, and if the effective instrument sensitivity
(after removing Galactic foregrounds) is significantly below
$5\,\mu\text{K}\,\text{arcmin}$ lensing can limit constraints on $r$.

We illustrate the impact of lensing on inflation constraints in
Table~\ref{tab:lens_inflation}.  We see that at a sensitivity of
$1\,\mu\text{K}\,\text{arcmin}$, the detection threshold for $r$ is
strongly limited by lensing. Fortunately, it is possible to remove
partially the lens-induced $B$-modes in a process known as
delensing~\citep{Kesden:2002ku,Knox:2002pe}.

\begin{table}
  \caption{\footnotesize Impact of lensing on constraints on the
    tensor-to-scalar ratio $r$ for a survey covering 70\,\% of the
    sky, with polarization sensitivity $1\,\mu\text{K}\,\text{arcmin}$
    (after foreground cleaning). For each model, the $1\sigma$ errors
    on $r$ are shown using only $B$-mode multipoles with $\ell<30$
    (i.e., the signal from reionization) and $\ell>30$ (the signal
    from recombination) and different assumptions about the level of
    delensing. The parameter $A_L$ describes the fraction of residual
    lensing $B$-mode power assumed after delensing, so that $A_L = 1$
    corresponds to no delensing, $A_L = 0$ to perfect delensing, and,
    for example, $A_L = 0.2$ to removing 80\,\% of the lensing power.}
\begin{center}
\begin{tabular}{@{\extracolsep{4pt}}ccccccccc@{}}
\hline
\hline
& \multicolumn{8}{c}{$10^4\times \sigma(r)$}  \\
\cline{2-9}
& \multicolumn{2}{c}{$A_L = 1$} & \multicolumn{2}{c}{$A_L = 0.5$}  & 
\multicolumn{2}{c}{$A_L = 0.2$}  & \multicolumn{2}{c}{$A_L = 0$} \\
\cline{2-3}\cline{4-5}\cline{6-7}\cline{8-9} 
Model & $\ell<30$ & $\ell>30$ & $\ell<30$ & $\ell>30$ & $\ell<30$ & $\ell>30$ & $\ell<30$ &
                                                                 $\ell>30$ \\
\hline
$r=0$ & 0.72 & 4.3 & 0.38 & 2.2 & 0.17 & 1.0 & 0.030 & 0.18 \\
$r=4\times 10^{-3}$ & 7.2 & 5.0 & 5.2 & 2.9 & 3.6 & 1.6 & 2.5 & 0.72 \\
$r=1\times 10^{-2}$ & 11 & 6.0 & 8.8 & 3.9 & 7.0 & 2.5 & 5.7 & 1.3 \\
\hline
\end{tabular}
\end{center}
\label{tab:lens_inflation}
\end{table}

Delensing of $B$-modes has recently been demonstrated in
practice~\citep{Planck2018VIII,Manzotti:2017net}, although with
current noise levels these analyses have not led to improved
inflationary constraints. However, for forthcoming ground-based
experiments delensing will be essential to exploit fully their
improved instrument sensitivities. These surveys will provide an
opportunity to refine our delensing algorithms to deal with real-world
issues such as variable depth observations, and to understand better
the interaction between Galactic foreground removal and
delensing. Biases, for example due to squeezed configurations of the
4-point function of polarized foregrounds, are currently poorly
understood due to a lack of high-quality data. Significant further
development of techniques will be required to achieve the necessary
accuracy of delensing for a space mission aiming for a level of $r$ of
$10^{-4}$. In addition, simulations of the turbulent, magnetized
interstellar medium, which faithfully capture the non-Gaussian
statistics over a sufficient dynamic range, can also play an important
role here.


\section{Statistical methods for large scale polarization}

A major challenge facing future polarization measurements of the CMB
is an accurate quantification of uncertainty in {\it cosmological
  parameters} in the presence of foregrounds, CMB lensing, and
instrumental systematics.  As stated before, current methods are not
reliable at the nanokelvin level required for $r \sim 10^{-4}$.  Many
different approaches have been proposed in the literature in order to
perform component separation \citep[see, e.g.,][]{Planck2013XII,
  Planck2015IX,Planck2018IV, Rocha2019}.  However, it is not clear
that these methods will suffice for future experiments, given the
uncertainties in the knowledge of the foreground emission and the
required nK sensitivity.  Within this context, tailored statistical
methods to analyse the large scale polarization signal are needed to
obtain a robust and unbiased constraint of the tensor-to-scalar ratio
$r$ and the reionization optical depth $\tau$ parameters.  Exampes
are:
(a) Gaussian Likelihood function computed exactly in pixel space
\citep{Gorski:1994, Slosar:2004fr, Page:2006hz, WMAP9}.  Although
optimal, this approach relies on the precise reconstruction of the
noise matrix in pixel space which can be extremely hard to achieve
when systematics, related to the instrument, the scanning strategy and
the residual foregrounds, dominate over noise.  New methods must
therefore be explored to solve this critical issue.
(b) The analysis of the \Planck\ HFI 100\,GHz and 143\,GHz large scale
$E$-modes polarization data to constrain
$\tau$ 
\citep{PlanckIntXLVI,PlanckIntXLVII} represented a first step in this
direction.
(c) A possible solution to the problem is given by defining the
likelihood in the harmonic space and using the CMB power spectra
calculated from the cross correlation of different frequencies and/or
data splits as input data \citep{2015MNRAS.453.3174M}, instead of maps
as in the pixel based likelihood method. This has the advantage of
greatly reducing the impact of uncorrelated residual systematics which
are different at each frequency and therefore do not bias the
cross-spectra.
(d) One important method for joint CMB and foreground reconstruction
is global Bayesian analysis. In this framework, the user must first
define a parametric model that accounts for cosmological,
astrophysical, and instrumental parameters.  The goal is then to map
out the full joint posterior distribution.  It is possible to sample
from this posterior distribution through Gibbs sampling.  Commander is
one specific implementation of this approach that was developed for,
and used extensively, by the \Planck\ collaboration.  At least three
conceptually different methods of constraining cosmological parameters
based on Gibbs sampling have been proposed in the literature so far:
the method adopted for the \Planck\ analysis called the Blackwell-Rao
estimator; and two other approaches
\citep[see][]{2013ApJ...777..150G,Racine2016}.  Further development of
these promising statistical approaches is crucial for future
experiments with nK sensitivity.

The simulator and analysis pipelines inherited from \Planck\ data
analysis represent great progress in accounting properly for the
foreground model uncertainties and instrument systematics. However,
the inevitable cross-talk between the foregrounds and the instrument
systematics, which can bias the estimation of cosmological parameters,
is not yet implemented there, and neither foreground nor instrument
systematics can yet be simulated or analyzed at the nanokelvin level.


\section{Concluding remarks}
We conclude by stating that significant work is required.  But we are
confident that with a few years of efforts, the requisite level of
accuracy can be achieved, and a space mission able to extract the full
range of information provided by the Universe at the nanokelvin level
can be designed, built, and flown.


\vspace{0.1in}
{\bf Acknowledgments} 
The research was carried out at the Jet Propulsion Laboratory, California Institute of Technology, under a contract with the National Aeronautics and Space Administration. 


\newpage
\bibliography{Kiss_WP_APC}
\bibliographystyle{astron}

\end{document}